\documentclass[twocolumn,preprintnumbers,amsmath,amssymb]{revtex4}

\usepackage{graphicx}
\usepackage{dcolumn}
\usepackage{bm}

\begin{document}


\title{Deformed dispersion relations and the Hanbury--Brown--Twiss Effect}
\author{Abel Camacho}
\email{acq@xanum.uam.mx}
\affiliation{Department of Physics \\
Universidad Aut\'onoma Metropolitana--Iztapalapa\\
Apartado Postal 55--534, C.P. 09340, M\'exico, D.F., M\'exico.}
\date{\today}

\begin{abstract}
In the present work we analyze the possibility of detecting some deformed dispersion relations, emerging in
some quantum--gravity models, resorting to the so--called Hanbury--Brown--Twiss effect.
It will be proved that in some scenarios the possibilities are not pessimistic. Forsooth, for some values
of the corresponding parameters the aforementioned effect could render interesting outcomes.
\end{abstract}
\maketitle

\section{Introduction}

The possibility that Lorentz symmetry could depict only an
approximate symmetry of quantum space has already been taken
seriously, and lot of work has been devoted to this analysis
\cite{[1], [2], [3]}. Within this realm several cases have been
studied, for instance, quantum--gravity approaches based upon
non--commutative geometry \cite{[4], [5]}, or loop--quantum
gravity models \cite{[6], [7]}, etc. Though the question of the
detection of these effects has always been considered an
impossible task, recently, \cite{[1], [3]}, this issue has been
addressed with a more optimisitic spirit.

One of the predicted traits, emerging in these a\-pproaches,
embodies a modified dispersion relation \cite{[3]}, the one
renders a small energy--dependent speed for the photon. The
feasibility around the detection of these co\-rrections has
already been analyzed, though it must be underlined that in the
aforementioned cases, it seems that the corresponding experiments
have always been consi\-dered in the realm of the so--called
first--order coherence properties of light \cite{[3], [8]}. Though
this sort of experiments have already played a fundamental role in
the context of gra\-vitational physics \cite{[9]}, it must be also
pointed out that they are very sensitive to vibrations and
fluctuations in the relative phase of the two involved waves, for
instance, due to the pro\-pagation through the atmosphere, and
hence, they could not be very useful in the present situation,
since we expect very tiny modifications.

In this spirit, in the present work the possibilities that the
so--called Hanbury--Brown--Twiss (HBT) effect could open up in
this realm are analyzed. The intention in our work is twofold.
Forsooth, firstly, as it is already known, HBT requires the use of
two photodetectors located at different points \cite{[8]}, and
concurrent with this last factor, it is not sensitive to
vibrations or atmospheric distortions, and in consequence it seems
to fit better with our goals. Secondly, this approach allows us to
introduce an additional parameter (the separation between the two
detectors), which could help us with our attempt of confronting
against the experiment some of these modified dispersion
relations. Indeed, as it will be shown below, the measurement of
this type of traits will depend (in a second--order coherence
process) not only upon the order of magnitude of the corrections,
but also upon the distance between the two involved
photodetectors. It will be proved that for some of the proposed
modifications the current technology could make possible the
detection of the corresponding new extra terms. In some other more
stringent scenarios, it will shown that a distance of $10^2$km.,
between the photodetectors, could render an interesting situation
in the experimental realm for the case of gamma--ray bursts.

\section{Second--order coherence and some quantum--gravity effects}

 As already mentioned above several quantum--gravity models predict a modified
dispersion relation \cite{[3]}, which can be characterized from a phenomenological point of
view through corrections hinging upon Planck's length, i.e., $l_p$

\begin{equation}
E^2 = p^2\Bigl[1 - \alpha\Bigl(El_p\Bigr)^n\Bigr].
\label{Disprel1}
\end{equation}

Here $\alpha$ is a coefficient, usually of order 1 and whose
precise value depends upon the considered quantum--gravity model,
and $n$, the lowest power in Planck's length leading to a
non--vanishing contribution, is also model dependent. Casting
(\ref{Disprel1}) in ordinary units

\begin{equation}
E^2 = p^2c^2\Bigl[1 - \alpha\Bigl(E\sqrt{G/(c^5\hbar)}\Bigr)^n\Bigr].
\label{Disprel2}
\end{equation}

Recalling that

\begin{equation}
p =\hbar k.
\label{Mom1}
\end{equation}

It is readily seen that

\begin{equation}
k =\frac{E/(c\hbar)}{\Bigl[1 - \alpha\Bigl(E\sqrt{G/(c^5\hbar)}\Bigr)^n\Bigr]^{1/2}}.
\label{k1}
\end{equation}

Since we expect very tiny corrections, then the following expansion is justified

\begin{equation}
k =\frac{E}{c\hbar}\Bigl[1 +
\frac{\alpha}{2}\Bigl(E\sqrt{G/(c^5\hbar)}\Bigr)^n +
\frac{3}{8}\alpha^2\Bigl(E\sqrt{G/(c^5\hbar)}\Bigr)^{2n}+...\Bigr].
\label{k2}
\end{equation}

Let us now consider two photons propagating along the axis defined
by the unit vector $\hat{e}$, but with different energy. At this
point it must be mentioned that in order to avoid a more
complicated experimental situation (namely, the consequences of a
light source having a continuous frequency distribution, in
connection with the presence of a deformed dispersion relation,
have not yet been addressed) we introduce only two frequencies.

\begin{equation}
\vec{k} = k\hat{e},
\label{k3}
\end{equation}

\begin{equation}
\vec{k}' = k'\hat{e}.
\label{k4}
\end{equation}

Let us now consider the detection of these photons resorting to HBT \cite{[8]}. In other words, we have two
photodetectors
located at points $A_1$ and $A_2$, with position vectors, $\vec{r}_1$ and $\vec{r}_2$, respectively.

The difference between HBT and the first--order co\-rrelation
function lies in the fact that the former measures the square of
the modulus of the complex degree of coherence, whereas a
first--order correlation function approach measures also the phase
\cite{[10]}.

The second--order correlation function reads \cite{[8]}

\begin{equation}
G^{(2)}(\vec{r}_1, \vec{r}_2; t,t) = \mathcal{E}\Bigl(1 +
\cos\Bigl[(\vec{k}-\vec{k}')\cdot(\vec{r}_2-\vec{r}_1)\Bigr]\Bigr).
\label{scf1}
\end{equation}

Here $\mathcal{E}$ is a constant factor with dimensions of electric field. With our previous expressions,
and denoting by $\Delta\theta^{(n)}$ the phase difference for  $n$, we have that the
interference pattern reads (to second--order in $\Delta E$)

\begin{eqnarray}
\Delta\theta^{(n)} = \frac{l\Delta E}{c\hbar}\Bigl[\Bigl(1 + \frac{n+1}{2}\alpha
[E\sqrt{G/(c^5\hbar)}]^n  \nonumber\\
+\frac{3}{8}\alpha^2(2n+ 1)[E\sqrt{G/(c^5\hbar)}]^{2n}\Bigr)\nonumber\\
+ \frac{\Delta E}{E}\Bigl(\frac{n(n+1)}{4}
\alpha[E\sqrt{G/(c^5\hbar)}]^n \nonumber\\
+ \frac{3n(2n+1)}{8}\alpha^2[E\sqrt{G/(c^5\hbar)}]^{2n}
\Bigr)\Bigr]. \label{phase1}
\end{eqnarray}

Here we have assumed that $E = E' + \Delta E$, and in addition, $l= \hat{e}\cdot(\vec{r}_2-\vec{r}_1)$.
The analysis of the feasibility of
the detection of this kind of corrections depends upon the value of $n$ \cite{[3]},
at least in the context of a first--order correlation function, and in consequence, we will divide our
situation in the same manner, namely, to first order in $\Delta E$ we have (approximately) that

\begin{eqnarray}
\Delta\theta^{(1)} = \frac{l\Delta E}{c\hbar}[1 +
\alpha[E\sqrt{G/(c^5\hbar)}]\nonumber\\
\times\Bigl(1 + \frac{9}{8}\alpha[E\sqrt{G/(c^5\hbar)}]\Bigr)],
\label{phase2}
\end{eqnarray}

\begin{eqnarray}
\Delta\theta^{(2)} = \frac{l\Delta E}{c\hbar}[1 +
\frac{3}{2}\alpha[E\sqrt{G/(c^5\hbar)}]^2\nonumber\\
\times\Bigl(1 + \frac{5}{4}\alpha[E\sqrt{G/(c^5\hbar)}]^2\Bigr)],
\label{phase3}
\end{eqnarray}

\section{Conclusions}

Let us now address the issue of the feasibility of this kind of experiments.
In order to do this let us assume that $\Delta E = \frac{E}{\gamma}$, with $\gamma >1$.
The possibility of measuring the involved corrections hinges upon the fact that if,
$\Delta\theta^{(0)}$ and $\Delta\theta^{(exp)}$ denote the phase difference
in the case in which $\alpha =0$, and the experimental resolution, respectively, then
$\Delta\theta^{(n)}- \Delta\theta^{(0)}> \Delta\theta^{(exp)}$.

Let us now contemplate this issue from a different pers\-pective, namely, we seek the value of
$l^{(n)}$, that renders the detection of the corrections. Hence (to first order
in $E\sqrt{G/(c^5\hbar)}]$)

\begin{eqnarray}
l^{(1)} \geq\frac{c\hbar\gamma}{\alpha E^2}\sqrt{c^5\hbar/G}\Delta\theta^{(exp)},
\label{length1}
\end{eqnarray}

\begin{eqnarray}
l^{(2)} \geq\frac{2}{3}\frac{c^6\hbar^2\gamma}{\alpha GE^3}\Delta\theta^{(exp)}.
\label{length2}
\end{eqnarray}

 If we assume the following values for our parameters, $\Delta\theta^{(exp)}\sim 10^{-4}$
 \cite{[10]},
$\alpha\sim 1$, $\gamma\sim 10^{2}$, $E\sim 10^{12}$e--V \cite{[1], [2], [3]}, then

\begin{eqnarray}
l^{(1)}\geq 10^{-3} cm,
\label{length3}
\end{eqnarray}

\begin{eqnarray}
l^{(2)}\geq 10^{13} cm.
\label{length4}
\end{eqnarray}

The energy that has been considered has the order of magnitude of the
highest energy that nowadays can be produced in a laboratory \cite{[3]}.
These two last expressions mean that if the corrections to the dispersion relation
entail $n=1$, then a HBT type--like experiment with a distance between the photodetectors
greater than
$10^{-3}$cm could detect the extra term. In the remaining case, $n=2$, the required distance
implies the impossibility of detecting (with this energy) the correction.

In the context of first--order correlation functions \cite{[8]}
the possibility of detecting the case $n = 2$ is, currently,
completely an impossible task. Forsooth, the time difference in
the arrival between the two photons is given by $10^{-18}$s.
Nevertheless, our a\-pproach introduces an additional parameter,
and therefore, if we consider the case of an energy of $E\sim
10^{19}$e--V (which is tantamount to the energy that could be
involved in the observation of gamma--ray bursts), then

\begin{eqnarray}
l^{(2)}\geq 10^{5}m. \label{length5}
\end{eqnarray}

Summing up, our input data has been the modified dispersion
relation that emerges in some quantum--gravity scenarios and,
resorting to the so--called second--order correlation function, it
has been proved that the most favorable case ($n = 1$) could be
analyzed within the cu\-rrent technological sensitivity.
Additionally, it has been shown that a more difficult situation
($n = 2$) could be tested with gamma--ray bursts.

A fleeting glimpse at the literature
shows us, that up to now, this last case ($n =2$) has been considered very difficult to
attack, experimentally. In the present model the presence of our extra parameter ($l$)
allows us to get closer to its possible detection.

Let us now address the topic of the feasibility of the present
proposal. There is already some experimental evidence \cite{[11]}
which purports that the case ($n= 1$) should be discarded. In
other words, experimentally we must consider $n=2$ as, physically,
more relevant than $n=1$. Therefore, we will analyze the
feasibility in the context of $n=2$, which is a tougher situation,
experimentally, to handle than $n=1$. The experimental parameter
that should be measured is the normalized correlation coefficient
of the fluctuations in the photoelectric current outputs
\cite{[12]}, $C(l)$. The connection with difference in time of
arrival stems from HBT, namely, the squared modulus of the degree
of coherence function, $\gamma$, is proportional to the normalized
correlation function of the photocurrent fluctuations,
\cite{[12]}, namely,

\begin{eqnarray}
C(l) = \delta\vert\gamma(l)\vert^2. \label{HBT1}
\end{eqnarray}

The parameter $\delta$ is the average number of photoelectric
counts due to light of one polarization registered by the detector
in the corresponding correlation time. Experimentally, for thermal
sources of temperature below $10^{5}$ K, $\delta$ is always
smaller than $1$ \cite{[12]}. In order to enhance the effect,
i.e., to have a larger value of $\delta$, we may consider the fact
that the number of average photons, as a function of the involved
frequency, $\nu$, reads \cite{[12]}

\begin{eqnarray}
\delta = 2\xi(3)\frac{\nu^3}{c^2\pi^2}. \label{MPN1}
\end{eqnarray}

Here $\xi$ is the so--called Riemann zeta function. Clearly,
higher energy implies larger mean number of photons. Hence, for an
energy of $E\sim 10^{19}$e--V we expect a value of $\delta$ not as
small as in the case of $10^{5}$ K. In other words, the higher the
energy of the light beam, the larger the constant between $C(l)$
and $\gamma$ becomes. Of course, this last fact cannot be
considered a shortcoming of the proposal.

In terms of the photocurrents fluctuations at the two
photodetectors

\begin{eqnarray}
C(l) = \frac{<\Delta I_1(t)\Delta I_2(t)>}{(<[\Delta
I_1(t)]^2>)^{1/2}(<[\Delta I_2(t)]^2>)^{1/2}}. \label{HBT2}
\end{eqnarray}

The feasibility of detecting a deformed dispersion relation in
this context depends upon the aforementioned fluctuations. We may
find already in the extant literature some models that explain the
pulse width of a Gamma Ray Burst (GRB), in terms of the involved
energy \cite{[13]}, as a power law expression, at least in the
case in which the sources are observed as fireballs. Though the
aforementioned result is a model it implies that fluctuations in
energy shall be present in GRB, and in consequence, as they
impinge upon the corresponding photodetectors they entail current
fluctuations. In other words, we may find non--vanishing sources
for $C(l)$, a fact that sounds promising, and that tells us that
we shall resort to those GRB (considering that we perform this
experiment within the range of $E\sim 10^{19}$e--V) which show the
largest pulse width. In this case we expect to have a better
experimental situation, and in consequence we may assert that this
proposal is a feasible one.

\begin{acknowledgments}
This research was supported by CONACYT Grant 42191--F. The author
would like to thank A.A. Cuevas--Sosa and A. Mac\'{\i}as for
useful discussions and literature hints.
\end{acknowledgments}

\end{document}